\documentclass[pre, preprint,showpacs,superscriptaddress]{revtex4-1}
\usepackage{graphicx}
\usepackage{color}
\usepackage[usenames,dvipsnames]{xcolor}
\usepackage[colorlinks=true,citecolor=ForestGreen,linkcolor=RubineRed,urlcolor=ForestGreen]{hyperref}
\usepackage{verbatim}
\usepackage{yfonts}
\usepackage{ulem}
\makeatletter

\usepackage{mathptmx} 

\usepackage{amsmath}
\usepackage{amsfonts}
\usepackage{amssymb}
\usepackage{epsfig}
\usepackage{dsfont}
\usepackage{mathrsfs}
\usepackage{enumitem}
\usepackage{graphicx}
\begin{document}
\title{Deformed Jarzynski Equality}
\author{Jiawen Deng}
\affiliation{NUS Graduate School for Integrative Science and Engineering, Singapore 117597}
\affiliation{Department of Physics, National University of Singapore, Singapore 117546}
\author{Juan D. Jaramillo}
\affiliation{Department of Physics, National University of Singapore, Singapore 117546}
\author{Peter H\"anggi}
\affiliation{Institute of Physics, University of Augsburg, Universit\"{a}tsstra{\ss}{\it e 1, D-86135} Augsburg, Germany}
\affiliation{Department of Physics, National University of Singapore, Singapore 117546}
\affiliation{Center for Quantum Technologies, National University of Singapore, Singapore 117543}
\affiliation{Nanosystems Initiative Munich, Schellingstra{\ss}e 4, 80799 M\"{u}nchen, Germany}
\author{Jiangbin Gong}\email{phygj@nus.edu.sg}
\affiliation{NUS Graduate School for Integrative Science and Engineering, Singapore 117597}
\affiliation{Department of Physics, National University of Singapore, Singapore 117546}


%
%




\begin{abstract}
The well-known Jarzynski equality, often written in the form $e^{-\beta\Delta F}=\langle e^{-\beta W}\rangle$,
provides a non-equilibrium means to measure the free energy difference $\Delta F$ of a system at the same inverse temperature $\beta$
based on an ensemble average of non-equilibrium work $W$.
The accuracy of Jarzynski's measurement scheme was known to be determined by
the variance of exponential work, denoted as ${\rm var}\left(e^{-\beta W}\right)$.
However, it was recently found that ${\rm var}\left(e^{-\beta W}\right)$
can systematically diverge in both classical and quantum cases.  Such divergence will necessarily pose
a challenge in the applications of Jarzynski equality because it
may dramatically reduce the efficiency in determining $\Delta F$.
In this work, we present a deformed Jarzynski equality for both classical and quantum non-equilibrium statistics,
in efforts to reuse experimental data
that already suffers from a diverging ${\rm var}\left(e^{-\beta W}\right)$.  The main feature of
our deformed Jarzynski equality is that it connects free energies at different temperatures and
it may still work efficiently subject to a diverging  ${\rm var}\left(e^{-\beta W}\right)$.
The conditions for applying our deformed Jarzynski equality may be met in experimental and computational situations. If so, then there is no need to redesign experimental or simulation methods.
Furthermore, using the deformed Jarzynski equality, we  exemplify the distinct behaviors of classical and quantum
work fluctuations  for the case of a time-dependent driven harmonic oscillator dynamics and provide insights into the essential performance differences between
classical and quantum Jarzynski equalities.
\end{abstract}
\date{\today}
\maketitle









\section{Introduction}

Work fluctuation theorems constitute one key topic in modern non-equilibrium
statistical mechanics \cite{ColloquiumRevModPhys.83.771,Bochkov_nonlinear_fluctuations,JarzynskiPRL,Crooks_relation,Hanggi_other_QFT}.
Of particular interest here is the Jarzynski equality (JE) $e^{-\beta\Delta F}=\langle e^{-\beta W}\rangle$\cite{JarzynskiPRL}, which links free energy differences
$\Delta F$ at the same inverse temperature $\beta$ with the ensemble average of exponential work $\langle e^{-\beta W}\rangle$, where work $W$ here refers to the inclusive form \cite{ColloquiumRevModPhys.83.771}. The
JE holds regardless of the details of a specific work
protocol, so long as the initial and final system configurations are fixed.
JE has stimulated vast interests in theory and experiments because it gives a mean of
direct measurement of free energy
difference $\Delta F$ by a finite sampling of work values $W$.  For $N$ sampled work values, one obtains

\begin{equation}
e^{-\beta\Delta F}=\langle e^{-\beta W}\rangle\approx\sum_{i=1}^{N}\frac{e^{-\beta W_{i}}}{N}.\label{eq:finite_sampling}
\end{equation}
This measurement scenario has been verified in a number of experiments for both classical
and quantum systems \cite{An_trapped_ion,Attila_experiment_pulling,Serra_experiment_closed_quantum,Lutz_trapped_ion,Harris_experiment_01,Hummer_experiment,Alemany_experiment_01,Douarche_classical_experiments_01,Liphardt_classical_experiment_01,Blickle_classical_experiment01,Ognjen_experiment_numerical_spring}.

Note that JE involves the ensemble of an exponential function of $W$.  As such,
rare events with large and negative $W$ could dominate the sample
average \cite{Jarzynski_rare_events_01,Jarzynski_rare_events_02,Zuckerman_error_estimation,Dellago_error_estimation_01,Dellago_error_estimation_02,Hartmann_simulation,Daura_simulation&error}.
This motivated people to study the error of $\Delta F$ through ${\rm var}\left(e^{-\beta W}\right)$
and some important insights have been obtained.  For example,  assuming a given precision to be reached for $\Delta F$, the corresponding required number of work realizations,
$N$, can be estimated by the central limit theorem (CLT) from the variance of exponential work, i.e.,
${\rm var}\left(e^{-\beta W}\right)$. Intuitively,
a larger ${\rm var}\left(e^{-\beta W}\right)$ requires more realizations
to reach the same precision in predicting $\Delta F$.  Suppression of work fluctuations by some control mechanism is hence desirable before applying JE.  For example, in Refs.~\cite{Deng2013,Xiaowork2014} we studied some classical and quantum control scenarios in efforts to minimize ${\rm var}\left(e^{-\beta W}\right)$.

Somewhat surprisingly, we recently found the possibility
of obtaining a systematic divergence in ${\rm var}\left(e^{-\beta W}\right)$ in classical systems, as verified
computationally in simple models isolated from a bath \cite{Deng_classical_divergence}. This divergence has immediate implications
for the applicability of JE in measuring $\Delta F$, but was seldom mentioned previously
except in some general discussions  made in Ref.~\cite{Zuckerman_error&divergence} and a specific result
on a one-dimensional gas undergoing an adiabatic
work protocol \cite{Dellago_error_estimation_01}.  That the divergence in  ${\rm var}\left(e^{-\beta W}\right)$
is not accidental can be understood as follows. For systems where the principle
of minimal work fluctuations \cite{Gaoyang_minimal_fluctuation} applies, if an adiabatic protocol yields a diverging ${\rm var}\left(e^{-\beta W}\right)$, the same quantity is expected to
diverge as well with increasing non-adiabaticity in the work protocols.
We stress that
a divergent ${\rm var}\left(e^{-\beta W}\right)$ makes CLT no longer
applicable, and the converging rate of $\sum_{i=1}^{N}e^{-\beta W_{i}}/N$ towards $e^{-\beta\Delta F}$
with respect to increasing $N$ is not obvious.  In one particular class of models \cite{Deng_classical_divergence},
a generalized version of CLT indicates that the converging rate is much slower
than the conventional $N^{-1/2}$ scaling law. Instead, the error is found to scale as $N^{-\gamma}$,  with the scaling exponent $\gamma$
being arbitrarily close to zero in extremely non-adiabatic cases. The lesson learned is that the
average $\sum_{i=1}^{N}e^{-\beta W_{i}}/N$ from experiments or simulations may barely converge
to the expected value $\langle e^{-\beta \Delta F}\rangle$ as $N$ increases.

Our further study reveals even more severe divergence problems in
quantum systems isolated from a bath \cite{Juan_quantum_divergence}. This indicates that quantum effects can play a crucial role
in work fluctuations.  Quantum JE and classical JE can thus have much different domains for meaningful applications.
For example, a work protocol could still lead to divergence in quantum ${\rm var}\left(e^{-\beta W}\right)$ even if its classical counterpart
has a finite ${\rm var}\left(e^{-\beta W}\right)$. In particular,
as the temperature characterizing a quantum system initially at thermal equilibrium decreases,
quantum effects become more appreciable and
then ${\rm var}\left(e^{-\beta W}\right)$ tends to diverge purely due to nonclassical effects \cite{Juan_quantum_divergence}.
This finding should have an important impact on future experimental studies of the quantum JE.

Our early results hence motivate us to consider the following situation.   Suppose an experiment or a computer simulation
has been carried out, and the ensemble average $\sum_{i=1}^{N}e^{-\beta W_{i}}/N$ based on finite sampling does not seem to converge.
A further checking on the quantity ${\rm var}\left(e^{-\beta W}\right)$ hints that it is probably a diverging quantity as $N$ increases.
Given such a situation, is there
a scheme to reprocess the data to extract useful predictions about equilibrium properties of the system (throughout this study, we assume no bath is involved during the work protocol)?  The aim of this work is to give a partially positive answer to this question.  We do so by
deforming the definition of the physical work to some quantity tunable, which in turn then yields
a deformed JE. So long as the variance of the exponential function of the newly defined quantity can be suppressed to a finite value,
then the deformed JE will work effectively.  We argue that our deformed JE proposed here is relevant to existing experiments \cite{An_trapped_ion,Attila_experiment_pulling,Serra_experiment_closed_quantum,Lutz_trapped_ion,Harris_experiment_01,Hummer_experiment,Alemany_experiment_01,Douarche_classical_experiments_01,Liphardt_classical_experiment_01,Blickle_classical_experiment01,Ognjen_experiment_numerical_spring}
and computational methods \cite{Daura_simulation&error,Dellago_computational01,Chipot_computational_01,Latorre_computational_01} motivated by JE.
As learned from our model studies, the deformed JE can eliminate the above-mentioned divergence issue
in classical cases effectively, but may not work well in the deep quantum regime. This observation itself also
exposes again the intrinsic difference between classical
and quantum JEs in terms of their potential applications.

This paper is organized as follows. We propose in Sec.2 a deformed classical JE. The model for illustration is a simple classical parametric harmonic oscillator (because such a model already suffices to show the divergence in ${\rm var}\left(e^{-\beta W}\right)$). In Sec. 3 we present a parallel deformed JE in the quantum domain and discuss why its usefulness is different from the classical version. Sec. 4 concludes this paper.
Throughout, our notation follows closely our early studies \cite{Deng_classical_divergence,Juan_quantum_divergence}.

\section{Classical Deformed JE}

\subsection{General Discussion}

We consider a general closed system whose Hamiltonian is given by $H(\mathbf{p},\mathbf{q};\lambda)$,
with phase space coordinates $(\mathbf{p},\mathbf{q})\in\Gamma$ and
a time-dependent control parameter $\lambda=\lambda(t)$.  The protocol starts
at $t=0$ and ends at $t=\tau$, i.e. $t\in[0,\tau]$. Given an initial
phase space coordinate $(\mathbf{p}_{0},\mathbf{q}_{0})$ as well as an arbitrary
work protocol $\lambda(0)=\lambda_{0}\rightarrow\lambda(\tau)=\lambda_{\tau}$,
the inclusive work \cite{ColloquiumRevModPhys.83.771} is obtained from

\begin{align}
W & =H(\mathbf{p}_{\tau},\mathbf{q}_{\tau};\lambda_{\tau})-H(\mathbf{p}_{0},\mathbf{q}_{0};\lambda_{0}),\label{eq:work_def_classical}
\end{align}
with $(\mathbf{p}_{\tau},\mathbf{q}_{\tau})=\big(\mathbf{p}(\mathbf{p}_{0},\mathbf{q}_{0},\tau),\mathbf{q}(\mathbf{p}_{0},\mathbf{q}_{0},\tau)\big)$
being the final phase space coordinate.
Note also that a proper gauge for the Hamiltonian $H(\mathbf{p},\mathbf{q};\lambda)$ is assumed here such that its value does equal to the energy of the system with no additional time-dependent gauge relying on $\lambda(t)$\cite{ColloquiumRevModPhys.83.771}.
Moreover (see below), we assume that this Hamiltonian is bounded from below while not being bounded from above. Let $Z(\lambda;\beta)=\int_{\Gamma}\exp\big(-\beta H(\mathbf{p},\mathbf{q};\lambda)\big){\rm d}\mathbf{p}{\rm d}\mathbf{q}$
be the partition function with parameter $\lambda$ and inverse temperature
$\beta$, which we assume to exist with $\beta>0$. Then the JE, which is valid for any protocol $\lambda_{0}\rightarrow\lambda_{\tau}$,
assumes the following form

\begin{align}
\langle e^{-\beta W}\rangle & =\int_{\Gamma}\exp\bigg[-\beta\left(H(\mathbf{p}_{\tau},\mathbf{q}_{\tau};\lambda_{\tau})-H(\mathbf{p}_{0},\mathbf{q}_{0};\lambda_{0})\right)\bigg]P_{0}(\mathbf{p}_{0},\mathbf{q}_{0};\lambda_{0};\beta){\rm d}\mathbf{p}_{0}{\rm d}\mathbf{q}_{0}\label{eq:average_classical}\\
 & =\frac{Z(\lambda_{\tau};\beta)}{Z(\lambda_{0};\beta)}=e^{-\beta\Delta F}\ .\label{eq:JE_classical}
\end{align}
Here $\langle\bullet\rangle$
represents taking average w.r.t.~the initial Gibbs distribution
$P_{0}(\mathbf{p},\mathbf{q};\lambda_{0};\beta)=\exp\big(-\beta H(\mathbf{p},\mathbf{q};\lambda_{0})\big)/Z(\lambda_{0};\beta)$.  We also assume that this ensemble average itself does exist (which may not be always the case \cite{Talkner_grand_canonical_divergent_Z}.
From the JE above, it is seen that $\langle e^{-\beta W}\rangle $ over the initial Gibbs state (at inverse temperature $\beta$)  will yield the same
$e^{-\beta\Delta F}$ once $H(\mathbf{p},\mathbf{q};\lambda_{0})$ and
$H(\mathbf{p},\mathbf{q};\lambda_{\tau})$ are fixed.
It is also clear that, the variance of exponential work, namely,  ${\rm var}\left(e^{-\beta W}\right)=\langle e^{-2\beta W}\rangle-e^{-2\beta\Delta F}$,
can be readily obtained through the second moment; i.e.,

\begin{align}
\langle e^{-2\beta W}\rangle & =\int_{\Gamma}\exp\bigg[-2\beta\big(H(\mathbf{p}_{\tau},\mathbf{q}_{\tau};\lambda_{\tau})-H(\mathbf{p}_{0},\mathbf{q}_{0};\lambda_{0})\big)\bigg]P_{0}(\mathbf{p}_{0},\mathbf{q}_{0};\lambda_{0};\beta){\rm d}\mathbf{p}_{0}{\rm d}\mathbf{q}_{0} \nonumber \\
 & =\int_{\Gamma}\frac{1}{Z(\lambda_{0};\beta)}\exp\bigg[-\beta\big(2H(\mathbf{p}_{\tau},\mathbf{q}_{\tau};\lambda_{\tau})-H(\mathbf{p}_{0},\mathbf{q}_{0};\lambda_{0})\big)\bigg]{\rm d}\mathbf{p}_{0}{\rm d}\mathbf{q}_{0}\ . \label{eq5}
\end{align}
To better illustrate the relation between classical and quantum cases,
we choose to use the adiabatic invariant $\Omega(E,\lambda)$ \cite{USSR,Brown_adiabatic_invariant_01,Brown_adiabatic_invariant_02,Edward_aidabatic_invariants_01,kasuga1961_1,kasuga1961_2,kasuga1961_3}
as an analogue of a quantum number indexing quantum energy levels. The adiabatic invariant is defined
as the phase space volume up to an energy $E$

\begin{equation}
\Omega(E;\lambda)=\int_{\Gamma}\Theta\big(E-H({\bf p},{\bf q};\lambda)\big){\rm d}{\bf p}{\rm d}{\bf q}\;,\label{eq:define_Omega}
\end{equation}
where $\Theta$ denotes the step function. Given that $\Omega(E, \lambda)$ equals the positive valued integrated density of states, $\Omega(E, \lambda)$ is monotonically growing with increasing energy $E$ \cite{footnote1}; therefore, the inverse function of $\Omega(E;\lambda)$
could be found from $E(\Omega;\lambda)$ in principle. Eqn.~(\ref{eq:average_classical})
could be rewritten as

\begin{equation}
\langle e^{-\beta W}\rangle=\int_{0}^{\infty}\int_{0}^{\infty}\exp\bigg[-\beta\big(E(\Omega_{\tau};\lambda_{\tau})-E(\Omega_{0};\lambda_{0})\big)\bigg]P(\Omega_{\tau}|\Omega_{0})P_{0}(\Omega_{0};\lambda_{0};\beta){\rm d}\Omega_{0}{\rm d}\Omega_{\tau}\ ,\label{eq:transition_classical}
\end{equation}
where $\Omega_{0}$ and $\Omega_{\tau}$ act like the initial and final
``energy index" while $P(\Omega_{\tau}|\Omega_{0})$ is the transition
probability between the states under a certain protocol, which is defined
in \cite{Deng_classical_divergence} and is proven to be bi-stochastic. Note here that the lower bounds $\Omega_0,\ \Omega_{\tau} = 0$ correspond to minimum of the lower bounded energy under fixed $\lambda_0$, $\lambda_\tau$. Likewise, Eqn.~(\ref{eq5}) can now be written as

\begin{equation}
\langle e^{-2\beta W}\rangle=\int_{0}^{\infty}\int_{0}^{\infty}\exp\bigg[-\beta\big(2E(\Omega_{\tau};\lambda_{\tau})-E(\Omega_{0};\lambda_{0})\big)\bigg]P(\Omega_{\tau}|\Omega_{0})P_{0}(\Omega_{0};\lambda_{0};\beta){\rm d}\Omega_{0}{\rm d}\Omega_{\tau}\ .\label{eq:transition_classical2}
\end{equation}

One can note from Eq.~(\ref{eq:transition_classical2}) that $\langle e^{-2\beta W}\rangle$ (and hence ${\rm var}\left(e^{-\beta W}\right)$)
diverges if, for example, we have $E(\Omega_{\tau};\lambda_{\tau})<E(\Omega_{0};\lambda_{0})/2$ for all $\Omega_\tau$.  To see this possibility clearly,
consider an adiabatic work protocol applied to a system with a constant density of states, with $\Omega(E,\lambda)\sim \lambda E$.
Then, because $\Omega(E,\lambda)$ must be an adiabatic invariant, we have $\lambda_{\tau} E(\Omega_{\tau}; \lambda_{\tau}) =\lambda_{0}E(\Omega_{0};\lambda_0)$, or $E(\Omega_{\tau};\lambda_{\tau})= (\lambda_{0}/\lambda_\tau)E(\Omega_{0};\lambda_0)$.  Thus, if $(\lambda_{0}/\lambda_\tau)<1/2$,  $\langle e^{-2\beta W}\rangle$ diverges. One can then infer \cite{Gaoyang_minimal_fluctuation} that any nonadiabatic protocol under fixed $\lambda_0$ and $\lambda_\tau$ will also yield a diverging ${\rm var}\left(e^{-\beta W}\right)$.

In order to overcome the divergence issue illustrated above, we now propose a deformed version
of JE for $H(\mathbf{p},\mathbf{q};\lambda)$ not bounded from above. The main idea is to
treat the statistics of an exponential function of $W_g$ (with $W_g=W$ if $g=1$) as a deformed $W$.
Specifically,  for an arbitrary value $g\in(0,1]$, we define $W_{g}$ as follows:

\begin{align}
W_{g} & =\frac{H(\mathbf{p}_{\tau},\mathbf{q}_{\tau};\lambda_{\tau})}{g}-H(\mathbf{p}_{0},\mathbf{q}_{0};\lambda_{0})\nonumber \\
 & =\frac{E(\Omega_{\tau};\lambda_{\tau})}{g}-E(\Omega_{0};\lambda_{0})\ .\label{eq:Wg_def}
\end{align}
The motivation to introduce the $g$-factor is to make the quantity $W_g$ less negative as compared with $W$ itself when it applies to transitions from high-energy initial states to low-energy final states.  Then, for positive $\beta$ values (which is assumed throughout this study)\cite{Hanggi_temperature_philosophy}, the exponential function $e^{-\beta W_{g}}$ would yield less dominating rare events and as a result, we hope that the variance in $e^{-\beta W_{g}}$ could be finite even when ${\rm var}\left(e^{-\beta W}\right)$ diverges \cite{footnote2}.

Consider then the ensemble average of $e^{-\beta W_{g}}$  over the same initial Gibbs state as used in the standard JE. We have

\begin{align}
\langle e^{-\beta W_{g}}\rangle & =\int_{0}^{\infty}\int_{0}^{\infty}\frac{1}{Z(\lambda_{0};\beta)}\exp\big(-\beta\frac{E(\Omega_{\tau};\lambda_{\tau})}{g}\big)P(\Omega_{\tau}|\Omega_{0}){\rm d} \Omega_{0}{\rm d}\Omega_{\tau}=\frac{Z(\lambda_{\tau};\beta/g)}{Z(\lambda_{0};\beta)}\  \label{dJE}.
\end{align}
In obtaning the second equality above we have used the bi-stochastic nature of $P(\Omega_{\tau}|\Omega_{0})$.
Equation (\ref{dJE}) indicates the following useful deformed JE:

\begin{equation}
F(\lambda_{\tau};\beta/g)-gF(\lambda_{0};\beta)=-\frac{g}{\beta}\ln\langle e^{-\beta W_{g}}\rangle\ . \label{dJE2}
\end{equation}
As seen from above, by calculating $\langle e^{-\beta W_{g}}\rangle$, we do not directly arrive at a free energy difference at the same inverse temperature $\beta$. Rather,  we would obtain a class of relations between the free energy $F(\lambda_{0};\beta)$ at the inverse temperature $\beta$  and a free energy $F(\lambda_{\tau};\beta/g)$ at inverse temperature $\beta/g$. This result for the special case $g=1$ recovers the original JE.
The potential benefit is that the second moment of $e^{-\beta W_{g}}$, i.e.,

\begin{equation}
\langle e^{-2\beta W_{g}}\rangle=\int_{0}^{\infty}\int_{0}^{\infty}\frac{1}{Z(\lambda_{0};\beta)}\exp\bigg(-\beta\big(2\frac{E(\Omega_{\tau};\lambda_{\tau})}{g}-E(\Omega_{0};\lambda_{0})\big)\bigg)P(\Omega_{\tau}|\Omega_{0}){\rm d}\Omega_{0}{\rm d}\Omega_{\tau}
\end{equation}
can be finite for a range of $g$ values even if it diverges for $g=1$.

In a typical classical experiment setup, the inclusive work $W$ is usually
measured along the work protocol in various ways \cite{Douarche_classical_experiments_01,Liphardt_classical_experiment_01,Blickle_classical_experiment01,Ognjen_experiment_numerical_spring}.
So is it feasible that $W_g$ defined in Eqn.~(\ref{eq:Wg_def})  can be indirectly measured along a classical trajectory? The answer is yes under certain conditions.  We suppose that in an experiment each individual value of $W$ is already measured. Then we may calculate
$W_{g}$ as

\begin{equation}
W_{g}=\frac{W}{g}+\frac{(1-g)}{g}E(\Omega_{0};\lambda_{0})\, \label{wgc}
\end{equation}
if additionally the initial energy $E(\Omega_{0};\lambda_{0})$ of each trajectory can be measured \cite{footnote3}.
That is, our knowledge of the initial energy values is the additional cost we have to pay in order to process our experimental data
through $W_g$ and $e^{-\beta W_{g}}$.  In addition, since all the initial energy values sampled from the Gibbs state are known, it is natural to assume that $F(\lambda_{0};\beta)$ is already known. Then from Eq.~(\ref{dJE2}) we can obtain $F(\lambda_{\tau};\beta/g)$.
Under these assumptions, we barely need to change an experiment setup to make use of our deformed JE.
Note also that in non-equilibrium numerical simulations \cite{Dellago_computational01,Zuckerman_error&divergence,Zuckerman_error_estimation},
the situation is even more obvious, because all the initial energy values of each sampling trajectory are by default registered in the simulations.

Let us now outline how to actually use our deformed JE for free energy measurements.  We assume that the aim is to measure the free energy $F(\lambda_{\tau};\bar{\beta})$ with $\bar{\beta}$ being the target inverse temperature.  As discussed above, this task may not be easily solved by use of JE because of the divergence in the second moment of exponential work.   We hence first
choose a trial $g$. Then, before applying a non-equilibrium work protocol, we prepare the system at thermal equilibrium at the inverse temperature $\beta=\bar{\beta}g$. Finally, we use the relation

\begin{equation}
F(\lambda_{\tau};\bar{\beta})=gF(\lambda_{0};\beta)-\frac{1}{\beta}\ln\langle e^{-\beta W_{g}}\rangle\
\end{equation}
to obtain $F(\lambda_{\tau};\bar{\beta})$.  We stress that the above procedure is mainly about a new way of reprocessing
the experimental or simulation data, with experimental or simulation details untouched.
Regarding the $g$ value to be determined, it can be in a range of values so long as the second moment of $e^{-\beta W_{g}}$
is not too large. This would be most appreciated, when JE suffers from the efficiency issue due to
a diverging $\langle e^{-2\beta W}\rangle$.  Next we illustrate the method by using the classical harmonic oscillator as an example
to show how a proper choice of $g$ indeed eliminates divergence in $\langle e^{-2\beta W_{g}}\rangle$.

\subsection{Classical Harmonic Oscillator}

We investigate a 1-dimensional (1D) harmonic oscillator with angular frequency $\omega>0$

\begin{equation}
H(p,q;\omega)=\frac{p^{2}}{2m}+\frac{1}{2}m\omega^{2}q^{2}.
\end{equation}
The equilibrium partition function for this system is given by $Z(\omega;\beta)=2\pi/\omega\beta$.
The phase space volume is given by

\begin{align}
\Omega(E;\omega) & =\int_{\Gamma}\Theta\big(E-\frac{p^{2}}{2m}-\frac{1}{2}m\omega^{2}q^{2}\big){\rm d}p{\rm d}q\nonumber \\
 & =2\pi\frac{E}{\omega}\ .
\end{align}
and $E(\Omega;\omega)=\omega\Omega/2\pi$. Clearly, this system belongs to the class of systems with a constant density of states
we used earlier for discussions. Work is done to the system as $\omega$ is forced to change with time, from $\omega_0$ to $\omega_\tau$. Under an arbitrary time-dependent $\omega(t),\ t\in [0,\tau]$, the transition probability $P(\Omega_{\tau}|\Omega_{0})$ can be calculated analytically. Particularly, the transition probability
under adiabatic driving is known to be $P(\Omega_{\tau}|\Omega_{0})=\delta(\Omega_{\tau}-\Omega_{0})$
\cite{Deng_classical_divergence}, thus by choosing $2\omega_{\tau}<\omega_{0}$,

\begin{align}
\langle e^{-2\beta W}\rangle_{{\rm ad}} & =\int_{0}^{\infty}\int_{0}^{\infty}\frac{1}{Z(\omega_{0};\beta)}\exp\bigg(-\beta\big(2E(\Omega_{\tau};\omega_{\tau})-E(\Omega_{0};\omega_{0})\big)\bigg)P(\Omega_{\tau}|\Omega_{0}){\rm d}\Omega_{0}{\rm d}\Omega_{\tau}\nonumber \\
 & =\int_{0}^{\infty}\frac{1}{Z(\omega_{0};\beta)}\exp\bigg(-\frac{\beta\Omega_{0}}{2\pi}\big(2\omega_{\tau}-\omega_{0}\big)\bigg){\rm d}\Omega_{0}=\infty\ .
\end{align}
Since adiabatic protocols minimize $\langle e^{-2\beta W}\rangle$,
we conclude all work protocols produce divergent $\langle e^{-2\beta W}\rangle$ as long as $2\omega_{\tau}<\omega_{0}$.

For an arbitrary non-adiabatic protocol, the transition probability can be calculated
explicitly for the harmonic oscillator (see Appendix \ref{appendix:HO_transition}) 
, yielding the following expression

\begin{equation}
P(\Omega_{\tau}|\Omega_{0})=\begin{cases}
\frac{1}{\pi\sqrt{(\Omega_{\tau}-\mu_{-}\Omega_{0})(\mu_{+}\Omega_{0}-\Omega_{\tau})}}\ , & \Omega_{\tau}\in[\mu_{-}\Omega_{0},\mu_{+}\Omega_{0}];\\
0\ , & {\rm otherwise.}
\end{cases}\label{eq:classical_transistion_prob}
\end{equation}
Here, $\mu_{\pm}$ are dimensionless constants satisfying $\mu_{+}\mu_{-}=1$
and $0<\mu_{-}<\mu_{+}$, which are determined merely by the protocol
$\lambda(t)$. Eqn.~(\ref{eq:classical_transistion_prob}) indicates that, given an initial $\Omega_{0}$,
the final $\Omega_{\tau}$ always fall in the interval $[\mu_{-}\Omega_{0},\mu_{+}\Omega_{0}]$.
One can also verify the bi-stochastic property, i.e.$\int_{0}^{\infty}P(\Omega_{\tau}|\Omega_{0}){\rm d}\Omega_{0}=\int_{0}^{\infty}P(\Omega_{\tau}|\Omega_{0}){\rm d}\Omega_{\tau}=1$.
With Eqn.~(\ref{eq:classical_transistion_prob}), the second
moment of $e^{-\beta W_g}$  can be found from

\begin{align}
\langle e^{-2\beta W_{g}}\rangle & =\int_{0}^{\infty}\int_{0}^{\infty}\frac{1}{Z(\omega_{0};\beta)}\exp\bigg(-\beta\big(2E(\Omega_{\tau};\omega_{\tau})-E(\Omega_{0};\omega_{0})\big)\bigg)P(\Omega_{\tau}|\Omega_{0}){\rm d}\Omega_{0}{\rm d}\Omega_{\tau}\nonumber \\
 & =\int_{0}^{\infty}\int_{\mu_{-}\Omega_{0}}^{\mu_{+}\Omega_{0}}\frac{1}{Z(\omega_{0};\beta)}\exp\bigg(-\frac{\beta}{2\pi}\big(2\frac{\omega_{\tau}\Omega_{\tau}}{g}-\omega_{0}\Omega_{0}\big)\bigg)P(\Omega_{\tau}|\Omega_{0}){\rm d}\Omega_{\tau}{\rm d}\Omega_{0}\nonumber \\
 & \le\int_{0}^{\infty}\int_{\mu_{-}\Omega_{0}}^{\mu_{+}\Omega_{0}}\frac{1}{Z(\omega_{0};\beta)}\exp\bigg(-\frac{\beta}{2\pi}\big(2\frac{\omega_{\tau}\mu_{-}\Omega_{0}}{g}-\omega_{0}\Omega_{0}\big)\bigg)P(\Omega_{\tau}|\Omega_{0}){\rm d}\Omega_{\tau}{\rm d}\Omega_{0}\nonumber \\
 & =\int_{0}^{\infty}\frac{1}{Z(\omega_{0};\beta)}\exp\bigg(-\frac{\beta\Omega_{0}}{2\pi}\big(2\frac{\omega_{\tau}\mu_{-}}{g}-\omega_{0}\big)\bigg){\rm d}\Omega_{0}\ .
\end{align}
The above inequality shows that if we choose $g<2\omega_{\tau}\mu_{-}/\omega_{0}$ (a sufficient but not necessary condition), then $\langle e^{-2\beta W_{g}}\rangle$ becomes finite. For such $g$ values, we can safely look into

\begin{equation}
W_{g}=\frac{E(\Omega_{\tau};\omega_{\tau})}{g}-E(\Omega_{0};\omega_{0}),
\end{equation}
whose second moment $\langle e^{-2\beta W_{g}}\rangle$ must be finite.  We have thus offered an explicit example where
the practical issue in applying JE due to a diverging second moment can be overcome by considering a deformed JE.

\section{Deformed Quantum JE}

\subsection{General discussion}

The inclusive work in quantum cases is obtained by
two-time measurements \cite{ColloquiumRevModPhys.83.771}, with

\begin{equation}
W=E_{j}^{\lambda_{\tau}}-E_{i}^{\lambda_{0}}\ ,\label{eq:work_def_quantum}
\end{equation}
where $E_{i}^{\lambda_{0}}$ and $E_{j}^{\lambda_{\tau}}$ are the
energies of projected eigenstates after measurements before and after
the work protocol. With the initial canonical distribution $P_{i}^{0}(\lambda_{0};\beta)$
and transition probability $P_{i\rightarrow j}$, the quantum JE can be obtained as follows:

\begin{equation}
\langle e^{-\beta W}\rangle=\sum_{i,j}\exp\big(-\beta(E_{j}^{\lambda_{\tau}}-E_{i}^{\lambda_{0}})\big)P_{i\rightarrow j}P_{i}^{0}(\lambda_{0};\beta)=\frac{Z(\lambda_{\tau};\beta)}{Z(\lambda_{0};\beta)},\label{eq:JE_quantum}
\end{equation}
where the bi-stochastic nature of $P_{i\rightarrow j}$, i.e.
$\sum_{i}P_{i\rightarrow j}=\sum_{j}P_{i\rightarrow j}=1$, has  been used.
The second moment in $\exp(-\beta W)$ is given by

\begin{equation}
\langle e^{-2\beta W}\rangle=\sum_{i,j}\text{\ensuremath{\frac{1}{Z(\lambda_{0};\beta)}}}\exp\big(-\beta(2E_{j}^{\lambda_{\tau}}-E_{i}^{\lambda_{0}})\big)P_{i\rightarrow j}\ .\label{eq:second_moment_quantum}
\end{equation}
As shown recently \cite{Juan_quantum_divergence}, this quantum second moment
can also diverge in systems with an infinite-dimensional Hilbert space. As a matter of fact,
the divergence in the quantum case occurs more frequently than in the classical case.

Driven by the same
motivation as outlined  in Sec.~2, we now define the corresponding quantum $W_{g}$ as

\begin{equation}
W_{g}=\frac{E_{j}^{\lambda_{\tau}}}{g}-E_{i}^{\lambda_{0}}\ . \label{wgc2}
\end{equation}
According to the two-time measurement  scheme of quantum work, the energy values of both the initial and final states need to be measured first.  Hence, there is no problem in obtaining $W_g$ from Eq.~(\ref{wgc2}), based on known values of $E_{j}^{\lambda_{\tau}}$ and $E_{i}^{\lambda_{0}}$.
One may then proceed to treat the ensemble average of $e^{-\beta W_{g}}$ and arrives at

\begin{eqnarray}
\langle e^{-\beta W_{g}}\rangle & = & \sum_{i,j}\exp\left[-\beta\left(\frac{E_{j}^{\lambda_{\tau}}}{g}-E_{i}^{\lambda_{0}}\right)\right]P_{i\rightarrow j}P_{i}^{0}(\lambda_{0};\beta)\nonumber \\
 & = & \sum_{i,j}\exp\left(-\beta\frac{E_{j}^{\lambda_{\tau}}}{g}\right)P_{i\rightarrow j}\frac{1}{Z(\lambda_{0};\beta)}\nonumber \\
 & = & \sum_{j}\exp\left(-\beta E_{j}^{\lambda_{\tau}}/g\right)\frac{1}{Z(\lambda_{0};\beta)}\nonumber \\
 & = & \frac{Z(\lambda_{\tau};\beta/g)}{Z(\lambda_{0};\beta)}\ .
\end{eqnarray}
This deformed quantum JE assumes precisely the same form as our previous classical result summarized by Eqn.~(\ref{dJE}).  Equation (\ref{dJE2}) hence also applies to the quantum case here.
The corresponding second moment in $e^{-\beta W_{g}}$ is given by

\begin{equation}
\langle e^{-2\beta W_{g}}\rangle=\sum_{i,j}\exp\left(-2\beta\bigg(\frac{E_{j}^{\lambda_{\tau}}}{g}-E_{i}^{\lambda_{0}}\bigg)\right)P_{i\rightarrow j}P_{i}^{0}(\lambda_{0};\beta)\ .\label{eq:Wg_second_moment_quantum}
\end{equation}
It is hoped that by also choosing  proper values of $g$, $\langle e^{-2\beta W_{g}}\rangle$  may merge as finite even when $\langle e^{-2\beta W}\rangle$ diverges. However, as we will show next, the situation in quantum cases can be much more challenging than in classical cases
due to some intrinsic differences between classical and quantum
state-to-state transition probabilities.

As a side note, in Appendix \ref{appendix:Deformed_Crooks} we have also presented a deformed quantum Crooks relation \cite{Crooks_relation} based on $W_g$. This will also help us understand better quantum deformed JE while motivating more interests in possible extensions of known fluctuation theorems.

\subsection{Quantum Harmonic Oscillator}

A closed quantum harmonic oscillator is described by the following Hamiltonian

\begin{equation}
\hat{H}(t)=\frac{\hat{p}^{2}}{2m}+\frac{1}{2}m\omega^{2}(t)\hat{q}^{2},
\end{equation}
where $\omega$ is still changing with time under a general work protocol.  For the quantum work statistics, it is convenient
to examine the so-called  characteristic function
\cite{Juan_quantum_divergence,Hanggi_work_is_not_observable,ColloquiumRevModPhys.83.771}
$G_{g}(\mu)$, which is the Fourier transformation of the probability distribution function $P(W_{g})$ for $W_{g}$. That is, we  have

\begin{equation}
G_{g}(\mu)=\int dW_{g}e^{\mathrm{i}\mu W_{g}}P(W_{g})\ .
\end{equation}
By choosing $\mu=2\mathrm{i}\beta$, we find (as a straightforward extension of the result for inclusive work $W$ in \cite{Juan_quantum_divergence})

\begin{align}
G_{g}(2\mathrm{i}\beta) & =\langle e^{-2\beta W_{g}}\rangle\nonumber \\
 & =\frac{\sqrt{2}{\sinh}(\beta\hbar\omega_{0}/2)}{\sqrt{{\cosh}\left(\beta\hbar(2\omega_{\tau}/g-\omega_{0})\right)-1-(Q^{\ast}-1){\sinh}(2\beta\hbar\omega_{\tau}/g){\sinh}(\beta\hbar\omega_{0})}}\ ,
\end{align}
where $Q^{\ast}$ is the Husimi coefficient determined solely by the
protocol $\omega(t)$ \cite{Husimi_harmonic}.  The deviation of $Q^{\ast}$ from unity describes the non-adiabaticity \cite{Husimi_harmonic} of a work protocol. Note that the case of $g=1$ reproduces our
previous result \cite{Juan_quantum_divergence} for the standard quantum work characteristic function.

Because our previous work provides sufficient details regarding the precise quantum-classical correspondence in terms of $P(W_{g=1})$ in the high temperature limit  \cite{Juan_quantum_divergence}, here we will not dive into the technical details regarding
the quantum-classical correspondence for $W_g$ with $g\ne 1$.  Instead, we just briefly mention that the quantum $P(W_{g})$ should also reduce to the corresponding classical distribution in the high temperature limit.

\begin{figure}[htbp]
\includegraphics[width=0.475\textwidth]{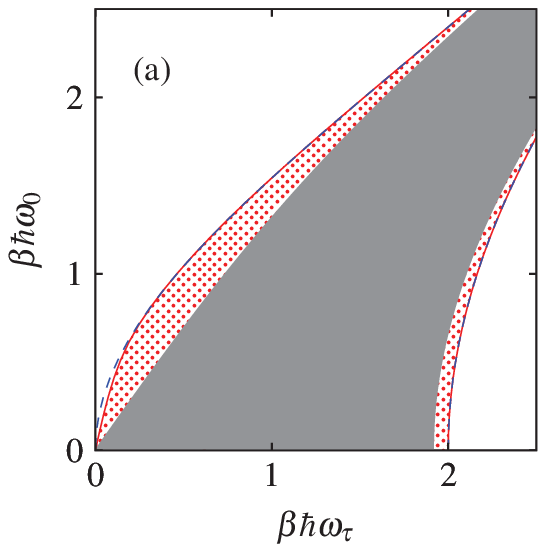}
\includegraphics[width=0.5\textwidth]{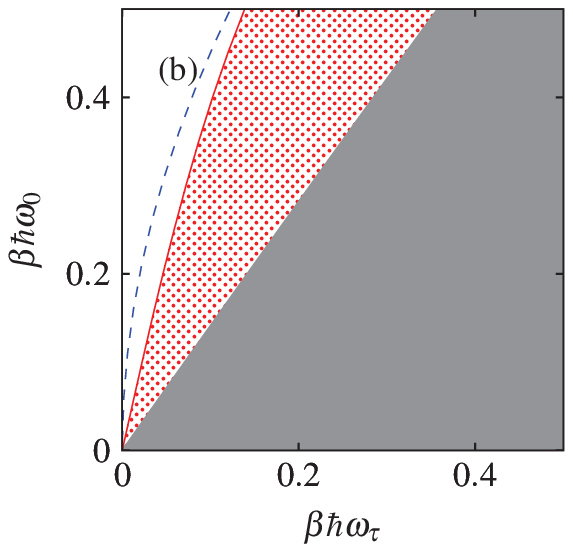}
\caption{Domains of convergence/divergence for $\langle e^{-2\beta W_{g}}\rangle$
under a sudden quench protocol from $\omega_0$ to $\omega_\tau$, which is applied to a quantum harmonic oscillator. The Husimi coefficient is given by $Q_{{\rm sq}}^{\ast}=\frac{1}{2}\left(\frac{\omega_{\tau}}{\omega_{0}}+\frac{\omega_{0}}{\omega_{\tau}}\right)$.
Here $\beta\hbar\omega_0$ and $\beta\hbar\omega_\tau$ denote the initial and final dimensionless angular frequencies $\beta\hbar\omega_0,\ \beta\hbar\omega_\tau$ of the oscillator  scaled by $\beta\hbar$, which is prepared initially at thermal equilibrium
 at the  inverse temperature $\beta$. For $g=1$, the second moment $\langle e^{-2\beta W_{g=1}}\rangle$
is finite only in the gray regime. For $g=0.1$ the domain for finite $\langle e^{-2\beta W_{g}}\rangle$
has grown to include the (red) patterned regimes as well. The (blue) dashed line is given by
Eqn.~\eqref{eq:quantum_divergence_criteria}, independent of $g$.  Note that this line almost exactly overlaps with the boundary of the (red) patterned domain in the lower right corner in (a).}
\label{fig:quantum_different_g}
\end{figure}

Remarkably though, the situation in the low temperature regime is much different. Figure 1 depicts the domain of finite $\langle e^{-2\beta W_{g}}\rangle $ for $g=1$ (grey area) and $g=0.1$ (grey area plus patterned area, with panel (b) focusing on cases with smaller values of $\beta\hbar\omega_0$ and $\beta\hbar\omega_\tau$ (hence more classical cases). Panel (b) clearly indicates that the use of $g=0.1$ has dramatically diminished the domain of divergence (white area) in $\langle e^{-2\beta W_{g}}\rangle $. However, if we examine panel (a) featuring more quantum regimes (large values of $\beta\hbar\omega_0$ and $\beta\hbar\omega_\tau$), it is seen that $g=0.1$ does not achieve much in suppressing the divergence domain.  We will explain this finding in the next subsection.

\subsection{Persistent Divergence of $\langle e^{-2\beta W_{g}}\rangle$ }

To gain analytical insights  we stick with the quantum harmonic oscillator case.   Without loss
of generality, we will focus on the even-parity states. The transition probabilities between even-parity states under an arbitrary  frequency-driving protocol are given by

\begin{equation}
P_{2i\rightarrow2j}=\bigg(\frac{2}{Q^{\star}+1}\bigg)^{1/2}\bigg(\frac{Q^{\star}-1}{Q^{\star}+1}\bigg)^{i+j}\frac{(2i)!(2j)!}{2^{2i+2j}}\bigg\{\sum_{m=0}^{\min\{i,j\}}\frac{[-2^{3}/(Q^{\star}-1)]^{m}}{(2m)!(j-m)!(i-m)!}\bigg\}^{2}\ \ .
\end{equation}
We next look into the asymptotic behaviour of the transition probabilities between
the initial $2i$th state to the final ground state ($j=0$) for very large
$i$:

\begin{align}
P_{2i\rightarrow0} & =\bigg(\frac{2}{Q^{\star}+1}\bigg)^{1/2}\bigg(\frac{Q^{\star}-1}{Q^{\star}+1}\bigg)^{i}\frac{(2i)!}{2^{2i}}\bigg(\frac{1}{i!}\bigg)^{2}\nonumber \\
 & \sim\bigg(\frac{2}{Q^{\star}+1}\bigg)^{1/2}\bigg(\frac{Q^{\star}-1}{Q^{\star}+1}\bigg)^{i}\frac{\sqrt{2\pi\times2i}(2i)^{2i}}{2^{2i}e^{2i}}\bigg(\frac{e^{i}}{i^{i}\sqrt{2\pi i}}\bigg)^{2}\nonumber \\
 & =\bigg(\frac{2}{Q^{\star}+1}\bigg)^{1/2}\bigg(\frac{Q^{\star}-1}{Q^{\star}+1}\bigg)^{i}\frac{1}{\sqrt{\pi i}}\ ,
\end{align}
where we have used  Stirling's formula $n!\sim\sqrt{2\pi n}n^{n}/e^{n}$ for very large $n$.

We are now ready to examine the contribution made by an individual transition $P_{2i\rightarrow0}$ to the second moment $\langle e^{-2\beta W}\rangle$ relevant to applications of the standard JE. It is found to be

\begin{align}
\exp\left(-2\beta\big(E_{0}^{\omega_{\tau}}-E_{i}^{\omega_{0}}\big)\right)P_{2i\rightarrow0}P_{i}^{0}(\lambda_{0};\beta) & =\frac{1}{Z(\lambda_{0};\beta)}\exp\left(-\beta\big(2E_{0}^{\omega_{\tau}}-E_{2i}^{\omega_{0}}\big)\right) {P}_{2i\rightarrow0}\\
 & =\frac{1}{Z(\lambda_{0};\beta)}\exp\left(-\beta\left(\hbar\omega_{\tau}-(2i+\frac{1}{2})\hbar\omega_{0}\right)\right){P}_{2i\rightarrow0}\nonumber \\
 & \sim\frac{\exp\left(-\beta\hbar(\omega_{\tau}-\frac{1}{2}\omega_{0})\right)}{Z(\lambda_{0};\beta)\ \sqrt{\pi i}}\bigg(\frac{2}{Q^{\star}+1}\bigg)^{1/2}\bigg(\frac{Q^{\star}-1}{Q^{\star}+1}e^{2\beta\hbar\omega_{0}}\bigg)^{i}\ . \label{eqg1}
\end{align}
Clearly, even this individual contribution diverges with increasing $i$ if the following condition is met:

\begin{equation}
\frac{Q^{\star}-1}{Q^{\star}+1}e^{2\beta\hbar\omega_{0}}>1\ .\label{eq:quantum_divergence_criteria}
\end{equation}
Dramatically, such a diverging contribution to $\langle e^{-2\beta W}\rangle$ has no classical analog. Indeed, according
to Eqn.~(\ref{eq:classical_transistion_prob}), the classical transition
probability from an initial highly excited state to a final low energy state is always strictly zero.  That is, it is the quantum non-vanishing transition probabilities that open ups a possibility for highly rare transitions to make contributions to exponential work fluctuations.  Put differently, the nonzero quantum transition probability $P_{2i\rightarrow0}$, though decreasing very fast with increasing $i$, can nevertheless make a diverging contribution to $\langle e^{-2\beta W}\rangle$ due to the exponential increasing factor arising from negative work values.  As a consequence, the more rare the initial state sampled from the Gibbs distribution is, the more it contributes to $\langle e^{-2\beta W}\rangle$.  This mechanism for quantum divergence in the second moment of exponential work can take effect, irrespective of whether or not the corresponding classical second moment of exponential work is finite.
This uncovers the potential difficulty in applying the standard quantum JE without first suppressing quantum work fluctuations \cite{Juan_quantum_divergence}.

Inspecting the parallel situation for $\langle e^{-2\beta W_{g}}\rangle$ reveals a similar problem.  The contribution made by the $P_{2i\rightarrow0}$ to $\langle e^{-2\beta W_{g}}\rangle$ is given by

\begin{align}
\exp\left(-2\beta\bigg(\frac{E_{0}^{\lambda_{\tau}}}{g}-E_{2i}^{\lambda_{0}}\bigg)\right)P_{2i\rightarrow0}P_{2i}^{0}(\lambda_{0};\beta) & \sim\frac{\exp\left(-\beta\hbar(\omega_{\tau}/g-\frac{1}{2}\omega_{0})\right)}{Z(\lambda_{0};\beta)\ \sqrt{\pi i}}\bigg(\frac{2}{Q^{\star}+1}\bigg)^{1/2}\bigg(\frac{Q^{\star}-1}{Q^{\star}+1}e^{2\beta\hbar\omega_{0}}\bigg)^{i}\ .
\end{align}
This expression is essentially the same as Eq.~(\ref{eqg1}), except for some  irrelevant factors.  Thus, we can again conclude that no matter what the $g$ value is, $\langle e^{-2\beta W_{g}}\rangle$ diverges under the condition of Eq.~(\ref{eq:quantum_divergence_criteria}).  This observation has important implications. Specifically, within the domain specified by Eq.~(\ref{eq:quantum_divergence_criteria}), the second moment $\langle e^{-2\beta W_g}\rangle$ always diverges, regardless of our choice of the $g$ values.  This theoretical insight is confirmed by our computational results in Fig. 1. In particular, from Fig. 1 it is seen that upon introducing $g=0.1$, the enlarged domain for finite $\langle e^{-2\beta W_{g}}\rangle$ cannot go beyond the dashed line.  Figure 1 in connection with the quantum divergence domain beyond the dashed line
also shows that, closer to the quantum regime (larger values of $\beta\hbar\omega_0$ and $\beta\hbar\omega_\tau$), most of the quantum divergence domain is occupied by the divergence domain determined by the above simple insight focusing on transitions from highly excited states to a single ground state. Thus, most of the quantum divergence domain shown in Fig. 1(a) cannot be removed by considering $W_g$. Given this insight, we note that closer to the classical regime,  the divergence in $\langle e^{-2\beta W_{g=1}}\rangle$ can occur outside the domain given by Eq.~(\ref{eq:quantum_divergence_criteria}) (for example, due to many other classical-like transitions). For the latter cases, $W_g$ is effective in removing divergences (see Fig. 1(b)).

\section{Conclusions}
Exponential work fluctuations characterized by ${\rm var}\left(e^{-\beta W}\right)$
or $\langle e^{-2\beta W}\rangle$ may systematically diverge \cite{Deng_classical_divergence,Juan_quantum_divergence}. This presents an
obstacle for a direct application of JE without effectively suppressing work fluctuations.
To meet this challenge in connecting non-equilibrium statistics with equilibrium properties,
we propose in this work a deformed work expression, denoted as $W_{g}$ and obtained deformed JE for both classical and quantum cases. This deformed
JE is based on an ensemble average of exponential quantities $e^{-\beta W_g}$ and connects this average with free energy values at different  temperatures.  Using the parametric harmonic oscillator as a test example,
we show that the classical deformed JE exhibits improved convergence features as compared to the case with the standard JE (with $g=1$),  because a possible divergent second moment $\langle e^{-2\beta W_g}\rangle$ can be be rendered convergent (i.e. finite) by a proper choice of $g\in(0,1]$.  This
tailored modification does not require a different design of simulation methods, but constitutes a beneficial possibility to reprocess the experimental data based on a finite number of
work realizations, yielding better performance.  As to the quantum deformed JE, it is shown that its performance may not be improved as effectively as compared with the standard quantum JE. This is because the
divergence in $\langle e^{-2\beta W_g}\rangle$ may not be lifted by introducing a  reduced positive $g$ value smaller than unity.
This feature reflects a fundamental difference between classical and quantum work statistics over exponential work functions.  While in classical cases, state-to-state transition probabilities can have very sharp cutoffs suppressing effectively the contributions from rare events; the state-to-state transition probabilities in quantum cases, though already exponentially suppressed, are not cut off sharply enough and
 can still create a scenario where more rare events make even larger contributions to ${\rm var}\left(e^{-\beta W}\right)$.   These findings indicate that
the efficiency of employing the quantum (standard or deformed) JE in predicting equilibrium properties is more limited than in the classical regime.
Given the insights gained from this study, it may  serve as an inspiration to seek other variants of deformed JEs and to apply those to different physical quantities that intrinsically make use of the conventional JE, classical or in its quantum form \cite{Allahverdyan_generalized_fluctuation,Deffner_generalized_fluctuation,Fusco_generalized_fluctuation,Hanggi_generalized_fluctuation,Talkner_generalized_fluctuation, Hanggi_quantum_work_PRE, Antonio_No-go_theorem, Deffner_entropy_open_system01}.

%




\bigbreak
\bigbreak

\acknowledgments{J.G. is supported by Singapore Ministry of Education Academic Research
Fund Tier-2 project (Project No. MOE2014-T2-2- 119 with WBS No. R-144-000-350-112).
P.H. also acknowledges support by the Singapore Ministry of Education and the National Research Foundation of Singapore.}

\appendix

\section{Transition Probabilities for Classical Harmonic Oscillator}
\label{appendix:HO_transition}
As discussed  in the main text, we consider the classical harmonic oscillator with a general time-dependent angular  frequency $\omega(t)>0$, whose Hamiltonian is given by

\begin{equation}
 H\big(p,q;\omega(t)\big)=\frac{p^{2}{2m}+\frac{1}{2}m\omega^{2}(t)q^{2}}\ .
\end{equation}
The phase space volume $\Omega(E;\omega)$ as defined in the main text is given by

\begin{align}
\Omega(E;\omega) & =\int_{\Gamma}\Theta\big(E-H(p,q;\omega)\big){\rm d}p{\rm d}q\nonumber \\
 & =\frac{2\pi E}{\omega}\ ,
\end{align}
with any $\omega(t)=\omega$ at fixed time $t$.
We define the transition probabilities as in Ref.~\cite{Deng_classical_divergence}:

\begin{equation}
P(\Omega_{\tau}|\Omega_{0})=\int_{\Gamma}\delta\bigg(\Omega_{\tau}-\Omega\big(H(p_{\tau},q_{\tau};\omega_{\tau});\omega_{\tau}\big)\bigg)\frac{\delta\big(E(\Omega_{0};\omega_{0})-H(p_{0},q_{0};\omega_{0})\big)}{\omega\big(E(\Omega_{0};\omega_{0});\omega_{0}\big)}{\rm d}p_{0}{\rm d}q_{0}\label{eq:transition_prob}
\end{equation}
where $(p_{\tau},q_{\tau})=\left(p(p_{0},q_{0},\tau),q(p_{0},q_{0},\tau)\right)$
denotes  the time evolution starting with $(p_{0},q_{0})$ and  ending at $t=\tau$, while $\omega\big(E(\Omega_{0};\omega_{0});\omega_{0}\big)$
represents the density of states. $\omega\big(E(\Omega_{0};\omega_{0});\omega_{0}\big)$  is also the normalization constant
for a micro-canonical ensemble at $E(\Omega_{0};\omega_{0})$, with

\begin{align}
\omega\big(E(\Omega_{0};\omega_{0});\omega_{0}\big) & =\int_{\Gamma}\delta\big(E(\Omega_{0};\omega_{0})-H(p_{0},q_{0};\omega_{0})\big){\rm d}p_{0}{\rm d}q_{0}\nonumber \\
 & =\frac{1}{\big(\frac{\partial E(\Omega;\omega)}{\partial\Omega}\big)_{\Omega_{0},\omega_{0}}}=\frac{2\pi}{\omega_{0}}\ .
\end{align}
Note that here there is no essential difference between using $\Omega$ and using $E$ for a harmonic
oscillator.  We however  still stick to using $\Omega$ since this notation is general.

To evaluate  Eqn.~(\ref{eq:transition_prob}) we examine the time evolution during $[0,\tau]$

\begin{align}
 q_{t }& =Yq_{0}+X\frac{p_{0}}{m}\\
 p_{t }& =m\dot{Y}q_{0}+\dot{X}p_{0}
\end{align}
where $X$, $Y$ represent  two special solutions satisfying  $X(0)=\dot{Y}(0)=0$ and $\dot{X}(0)=Y(0)=1$ \cite{Husimi_harmonic}. $X$, $Y$ are independent of $(p_{0},q_{0})$ due to the fact that the equations of motion are linear. Let

\begin{align*}
x & =\frac{p_{0}}{\sqrt{m\omega_{0}}}\ ; \\
y & =\sqrt{m\omega_{0}}q_{0}\ .
\end{align*}
Eqn.~(\ref{eq:transition_prob}) then becomes

\begin{equation}
P(\Omega_{\tau}|\Omega_{0})=\int_{\Gamma}\frac{\omega_{0}}{2\pi}\delta\big(\Omega_{\tau}-\pi\gamma^{T}M\gamma\big)\delta\big(\frac{\omega_{0}\Omega_{0}}{2\pi}-\frac{\omega_{0}}{2}(x^{2}+y^{2})\big){\rm d}\gamma\label{eq:transition_prob_intermediate}
\end{equation}
where $\gamma=(x,y)^{T}$ and $d\gamma=dxdy$, and

\begin{equation}
M=\bigg(\begin{array}{cc}
(X^{2}\omega_{0}\omega_{\tau}+\dot{X}^{2}\frac{\omega_{0}}{\omega_{\tau}}) & (\dot{X}\dot{Y}\frac{1}{\omega_{\tau}}+XY\omega_{\tau})\\
(\dot{X}\dot{Y}\frac{1}{\omega_{\tau}}+XY\omega_{\tau}) & (\dot{Y}^{2}\frac{1}{\omega_{0}\omega_{\tau}}+Y^{2}\frac{\omega_{\tau}}{\omega_{0}})
\end{array}\bigg)
\end{equation}
is a symmetric matrix. Diagonalizing $M$ yields

\begin{equation}
M=O^{T}DO,
\end{equation}
where

\begin{equation}
D=\bigg(\begin{array}{cc}
\mu_{+} & 0\\
0 & \mu_{-}
\end{array}\bigg)
\end{equation}
is diagonal with $\mu_{+}>\mu_{-}$ and $O$ is an orthogonal matrix. Note also that
 $M$ is positive definite, therefore $\mu_{+}>\mu_{-}>0$. One can further show that $\mu_+\mu_-=1$ by noticing that $X\dot{Y}-\dot{X}Y=1$ throughout the protocol. With these results,
Eq.~(\ref{eq:transition_prob_intermediate}) becomes

\[
P(\Omega_{\tau}|\Omega_{0})=\int_{\Gamma}\frac{1}{2\pi}\delta\big(\frac{\Omega_{0}}{2\pi}-\frac{1}{2}(x^{2}+y^{2})\big)\delta\bigg(\Omega_{\tau}-\pi(\mu_{+}x'^{2}+\mu_{-}y'^{2})\bigg){\rm d}\gamma'
\]
by letting $\gamma'=(x',y')^{T}=O\gamma$. We next replace $(x',y')$
by $(\rho\cos\theta,\rho\sin\theta)$ as a change of integration variables, we obtain

\begin{align}
P(\Omega_{\tau}|\Omega_{0}) & =\int_{\Gamma}\frac{1}{2\pi}\delta\big(\frac{\Omega_{0}}{2\pi}-\frac{1}{2}(x^{2}+y^{2})\big)\delta\big(\Omega_{\tau}-\pi\gamma^{T}M\gamma\big){\rm d}\gamma\nonumber \\
 & =\int_{0}^{2\pi}\int_{0}^{\infty}\frac{1}{2\pi}\delta\big(\frac{\Omega_{0}}{2\pi}-\frac{1}{2}\rho^{2}\big)\delta\bigg(\Omega_{\tau}-\pi(\mu_{+}x'^{2}+\mu_{-}y'^{2})\bigg)\rho {\rm d}\rho {\rm d}\theta\nonumber \\
 & =\int_{0}^{2\pi}\int_{0}^{\infty}\frac{1}{2\pi}\delta\big(\frac{\Omega_{0}}{2\pi}-\frac{1}{2}\rho^{2}\big)\delta\bigg(\Omega_{\tau}-\pi\rho^{2}(\mu_{+}\cos^{2}\theta+\mu_{-}\sin^{2}\theta)\bigg){\rm d}\frac{\rho^{2}}{2}{\rm d}\theta\nonumber \\
 & =\int_{0}^{2\pi}\frac{1}{2\pi}\delta\bigg(\Omega_{\tau}-\Omega_{0}\big((\mu_{+}-\mu_{-})\cos^{2}\theta+\mu_{-}\big)\bigg){\rm d}\theta
\end{align}
For $P(\Omega_{\tau}|\Omega_{0})\neq0$, we require
\[
\mu_{-}\Omega_{0}<\Omega_{\tau}<\mu_{+}\Omega_{0} \ .
\]
Under this condition we arrive at

\begin{equation}
P(\Omega_{\tau}|\Omega_{0})=\frac{1}{\pi\sqrt{(\Omega_{\tau}-\mu_{-}\Omega_{0})(\mu_{+}\Omega_{0}-\Omega_{\tau})}}\ . \label{eqn:CHO_transition_probability}
\end{equation}
One can then also verify the following bi-stochastic condition

\begin{equation}
\int_{0}^{\infty}P(\Omega_{\tau}|\Omega_{0}){\rm d}\Omega_{0}=\int_{0}^{\infty}P(\Omega_{\tau}|\Omega_{0}){\rm d}\Omega_{\tau}=1.
\end{equation}

\section{Deformed Crooks Relation}
\label{appendix:Deformed_Crooks}
Consider the \textit{forward} characteristic function of work,

\begin{equation}
G_{+,g}(\mu)=\int dW_{g}e^{\mathrm{i}\mu W_{g}}P(W_{g}),\label{crk1}
\end{equation}
where,

\begin{equation}
P_{+}(W_{g})=\sum_{i,j}\delta(W_{g}-E_{j}^{\lambda_{\tau}}/g+E_{i}^{\lambda_{0}})P_{i\rightarrow j}P_{i}^{0}(\lambda_{0};\beta).\label{crk2}
\end{equation}
Substituting Eq.~\eqref{crk2} into Eq.~\eqref{crk1},  one has

\begin{equation}
G_{+,g}(\mu)=\sum_{i,j}{\rm exp}[\mathrm{i}\mu(E_{j}^{\lambda_\tau}/g-E^{\lambda_0}_{i})]P_{i\rightarrow j}P_{i}^{0}(\lambda_{0};\beta).
\end{equation}
Further,

\begin{eqnarray}
G_{+,g}(\mu+\mathrm{i}\beta) & = & \sum_{i,j}{\rm exp}[\mathrm{i}\mu(E_{j}^{\lambda_{\tau}}/g-E_{i})]{\rm exp}[-\beta(E_{j}^{\lambda_{\tau}}/g-E_{i})]P_{i\rightarrow j}P_{i}^{0}(\lambda_{0};\beta)\nonumber \\
 & = & \frac{1}{Z(\lambda_{0};\beta)}\sum_{i,j}{\rm exp}[\mathrm{i}\mu(E_{j}^{\lambda_{\tau}}/g-E_{i}^{\lambda_{0}})]{\rm exp}(-\beta E_{j}^{\lambda_{\tau}}/g)P_{i\rightarrow j}\nonumber \\
 & = & \frac{Z(\lambda_{\tau};\beta/g)}{Z(\lambda_{0};\beta)}\sum_{i,j}{\rm exp}[\mathrm{i}\mu(E_{j}^{\lambda_{\tau}}/g-E_{i}^{\lambda_{0}})][{\rm exp}(-\beta E_{j}^{\lambda_{\tau}}/g)/Z(\lambda_{\tau};\beta/g)]P_{i\rightarrow j},
\end{eqnarray}
where in the last line we have simultaneously multiplied by $Z(\lambda_{\tau};\beta/g)=\sum_{j}{\rm exp}(-\beta E_{j}^{\lambda_{\tau}}/g)$.
Identifying the following Gibbs distribution at inverse temperature $\beta/g$ as

\begin{equation}
P_{g,j}^{\tau}={\rm exp}(-\beta E_{j}^{\lambda_{\tau}}/g)/Z(\lambda_{\tau};\beta/g),
\end{equation}
and using $P_{i\rightarrow j}=P_{j\rightarrow i}$,
we end up with

\begin{eqnarray}
G_{+,g}(\mu+\mathrm{i}\beta) & = & \frac{Z(\lambda_{\tau};\beta/g)}{Z(\lambda_{0};\beta)}\sum_{i,j}{\rm exp}[\mathrm{i}\mu(E_{j}^{\lambda_{\tau}}/g-E_{i}^{\lambda_{0}})]P_{j\rightarrow i}P_{g,j}^{\tau}\nonumber \\
 & = & \frac{Z(\lambda_{\tau};\beta/g)}{Z(\lambda_{0};\beta)}\int dW_{g}\sum_{i,j}{\rm exp}(-\mathrm{i}\mu W_{g})\delta(W_{g}-E_{i}^{\lambda_{0}}+E_{j}^{\lambda_{\tau}}/g)P_{j\rightarrow i}P_{g,j}^{\tau}.
\end{eqnarray}
Observing that

\begin{equation}
P_{-}(-W_{g})=\int dW_{g}\sum_{i,j}\delta(W_{g}-E_{i}^{\lambda_{0}}+E_{j}^{\lambda_{\tau}}/g)P_{j\rightarrow i}P_{g,j}^{\tau},
\end{equation}
we find

\begin{equation}
G_{+,g}(\mu+\mathrm{i}\beta)=\frac{Z(\lambda_{\tau};\beta/g)}{Z(\lambda_{0};\beta)}\sum_{i,j}{\rm exp}(-\mathrm{i}\mu W_{g})P_{-}(-W_{g}).
\end{equation}
Likewise we obtain

\begin{equation}
G_{-,g}(\mu)=\int dW_{g}e^{\mathrm{i}\mu W_{g}}P_{-}(W_{g}).
\end{equation}
Consequently we have the relation that

\begin{equation}
G_{+}(\mu+\mathrm{i}\beta)=\frac{Z(\lambda_{\tau};\beta/g)}{Z(\lambda_{0};\beta)}G_{-}(-\mu).\label{crk3}
\end{equation}
After Fourier transform in both sides, one obtains the Crooks relations.
Indeed, for the LHS of \eqref{crk3}

\begin{eqnarray}
\int d\mu e^{-\mu W'_{g}}G_{+}(\mu+\mathrm{i}\beta) & = & \int d\mu\int dW_{g}e^{-\mu(W'_{g}-W_{g})}e^{-\beta W_{g}}P_{+}(W_{g})\nonumber \\
 & = & \int dW_{g}\delta(W'_{g}-W_{g}){\rm exp}(-\beta W_{g})P_{+}(W_{g})\nonumber \\
 & = & {\rm exp}(-\beta W'_{g})P_{+}(W'_{g})
\end{eqnarray}
while for the RHS of \eqref{crk3},

\begin{eqnarray}
\int d\mu G_{-}(-\mu) & = & \int d\mu\int dW_{g}e^{-\mu(W'_{g}-W_{g})}P_{-}(-W_{g})\nonumber \\
 & = & P_{-}(-W'_{g})
\end{eqnarray}
Therefore, \eqref{crk3} can be written as

\begin{equation}
e^{-\beta W'_{g}}P_{+}(W'_{g})=\frac{Z(\lambda_{\tau};\beta/g)}{Z(\lambda_{0};\beta)}P_{-}(-W'_{g})\ .
\end{equation}
This result then coincides  with the standard Crooks relation for  $g\rightarrow1$.



\renewcommand\bibname{References}



\begin{thebibliography}{-------}
\bibitem{footnote1}
Here we assume that the density of states and thus $\Omega(E,\lambda)$ is not growing in energy exponentially, as otherwise the free energy would not excist at sufficiently large temperature $\beta \rightarrow 0$.
\bibitem{footnote2}
Heuristically, divergent $e^{-2\beta W}$ is a consequence of negative work $W<0$. Because we assume the system to be lower bounded but not upper bounded, then choosing $g\in (0,1]$ implies $E(\Omega_\tau;\lambda_\tau)/g>E(\Omega_\tau;\lambda_\tau)$
for large $\Omega_\tau$, therefore suppressing the emergence of negative work.
\bibitem{footnote3}
For example this is feasible for the class of  single particle Hamiltonians of known structure (e.g. set of harmonic oscillators) with their mutual  interaction being switched on via the control parameter $\lambda(t)$ only for times $t \geq 0^+$.
\end{thebibliography}


\end{document}